\documentclass{llncs}

\usepackage[english]{babel}
\usepackage{float}
\usepackage{graphicx,epstopdf}
\usepackage{amsmath,amssymb,amsfonts,subfigure}
\usepackage{comment}
\usepackage{cite}
\usepackage{adjustbox}

\title{Rank me thou shalln't Compare me}
\author{Akrati Saxena, Vaibhav Malik, S. R. S. Iyengar\\
\normalsize{\{akrati.saxena, vaibhav.malik, sudarshan\}@iitrpr.ac.in}\\
}
\institute{Department of Computer Science and Engineering,\\
\normalsize{Indian Institute of Technology Ropar, India}}

\begin{document}
\maketitle
\begin{abstract}

Centrality measures have been defined to quantify the importance of a node in complex networks. The relative importance of a node can be measured using its centrality rank based on the centrality value. In the present work, we predict the degree centrality rank of a node without having the entire network. The proposed method uses degree of the node and some network parameters to predict its rank. These network parameters include network size, minimum, maximum, and average degree of the network. These parameters are estimated using random walk sampling techniques. The proposed method is validated on Barabasi-Albert networks. Simulation results show that the proposed method predicts the rank of higher degree nodes with more accuracy. The average error in the rank prediction is approximately $0.16\%$ of the network size.
\end{abstract}

\section{Introduction}

Complex networks have attracted researchers from the past few years. Complex networks \cite{bar1997dynamics} have been the part of our day to day life such as friendship networks \cite{wasserman1994social}, collaboration networks \cite{newman2001structure}, World Wide Web \cite{albert1999internet}, Internet \cite{huberman1999internet}, biological networks \cite{hartwell1999molecular}, and so on. In complex networks, objects are represented by nodes and the relationship between a pair of nodes is represented by an edge connecting them. Researchers have been studying the evolving phenomenon and properties of these networks for a fairly long time. 

All real world complex networks are sparse and follow some properties such as small world phenomenon (six degrees of separation) \cite{milgram1967small}, scale-free degree distribution \cite{barabasi1999emergence}, preferential attachment \cite{jeong2003measuring}, high clustering coefficient \cite{klemm2002highly}, etc. Each property shows a different aspect of these networks. For example, small world phenomenon shows that no two nodes are far from each other. Real world networks have high clustering coefficient, that shows how tightly knit a node is in its neighborhood. In friendship networks, a user has high probability to be friend with friends of her friends because of high trust factor and frequent encounters. This gives birth to high clustering coefficient due to the increase in number of triangular links. The topological structure of complex networks also has meso-scale properties which include community structure \cite{clauset2004finding}, and core-periphery structure \cite{borgatti2000models, saxena2016evolving}.


Except these properties, each node also possess some unique characteristics. These properties can be measured using different centrality metrics. These centrality measures compute the importance of a node under different contexts. Some of these centrality measures can be computed using local information of the node like degree centrality \cite{shaw1954some}, and semi-local centrality measure \cite{chen2012identifying}. Others use global information of the network like  closeness centrality \cite{sabidussi1966centrality}, betweenness centrality \cite{freeman1977set}, eigenvector centrality \cite{stephenson1989rethinking}, katz centrality \cite{katz1953new}, pagerank \cite{brin1998anatomy}, and so on. The computation of global centrality measures is very costly as they require entire structure of the network.

These centrality measures assign an index value to each node. But in real life applications, most of the time actual value is not important, what's important is where you stand, not with respect to the mean but with respect to everyone else. For example, in most of the entrance exams, percentile of the candidate is considered for the shortlisting not the percentage, as percentile tells about the rank. Similarly in social networks, nodes might be interested in computing their rank based on different centrality measures. 

In the present work, we predict the global rank of a node based on its degree. The degree of a node denotes the total number of neighbors of the node. In social networks, the degree rank of a node denotes how popular or strong a node is in the given network. A node having more neighbors is stronger and has high rank. Similarly in WWW network, degree rank of a node denotes the relative importance of a particular article or topic that is associated with the web page. 

One simple way to compute the degree rank of a node is, collect the degree values of all nodes, and compare them to get the rank of the interested node. This method requires complete structure of the network in hand. But it is not feasible to collect and store the entire data on a single system, as the size of complex networks is increasing very fast with time. These networks are highly dynamic, and the collected dataset needs to be updated regularly. This has inspired researchers to propose approximation algorithms based on the local information. The local information of the node or the network can be collected using various sampling techniques. Authors have used random walk sampling techniques and its variations to estimate network parameters like network size \cite{hardiman2013estimating, kurant2012graph, katzir2011estimating, cooper2014estimating, goldreich2008approximating}, clustering coefficient \cite{hardiman2013estimating}, average degree \cite{dasgupta2014estimating}, online polls \cite{dasgupta2012social}, etc. These approaches are storage efficient as a small snapshot of the dataset is collected to process the request.

In this work, we propose a method to predict the degree rank of a node. The proposed method uses power law degree distribution characteristic of the network and few network parameters like network size, maximum, minimum and average degree of the network. These parameters are estimated using random walk sampling techniques in pre-processing steps. The network size is estimated using the method proposed by Hardiman and Katzir \cite{hardiman2013estimating}. The minimum and maximum degree is estimated as the minimum and maximum degree available in the sample. The average degree of the network is estimated using the method proposed by Dasgupta et al. \cite{dasgupta2014estimating}. 

The proposed method is simulated on BA networks to validate its accuracy. The error in the predicted rank is computed as the modular difference of the actual and the predicted rank. Simulation results show that the proposed method estimates the rank with high accuracy. We have also proposed a probabilistic method to estimate the degree rank of a node \cite{saxena2015estimating}. As per the best of our knowledge, this is the first work of its kind\footnote{This work has been published at \cite{saxena2016estimating}.}. The local estimation of the global rank can help to identify influential nodes in the network. Identification of influential nodes has been the core of various research problems like epidemic, memetics \cite{gupta2016modeling}, viral marketing, information diffusion \cite{saxena2015understanding}, opinion formation, and so on. The rest of this paper is organized as follows. Section 2 describes BA model followed by the mathematical analysis of degree ranking method. Section 3 contains experimental results and the paper is concluded in section 4.

\section{Model and Background}

In this section, we introduce the mathematical analysis to approximate degree rank of a node. We will use following notations throughout our discussion. Let's assume that the given graph $G$ has $n$ nodes. Minimum, maximum, and average degree of the network are represented by $k_{min}$, $k_{max}$, and $d_{avg}$ respectively. All these parameters are estimated using sampling methods. Degree of a node $u$ is denoted by $k_u$, that represents total number of neighbors of the node. Degree rank of a node $u$ is defined as, $R_a(u) = \sum_v X_{vu}+1$, where $X_{vu}$ is 1 when $deg(v) > deg(u)$ and 0 otherwise. It is also called actual degree rank of the node in the given network $G$. The predicted degree rank of a node $u$ is denoted by $R_p(u)$. Next, we discuss the preferential attachment model (BA model) to generate scale-free synthetic networks.

\subsection{BA Model}
Barabasi and Albert proposed an evolutionary preferential attachment model to generate synthetic networks that follows the properties of real world scale-free complex networks \cite{barabasi1999emergence}. This model starts with a seed graph that contains $n_0$ disconnected nodes. At each time stamp, a new node is added and it is connected with $m$ already existing nodes. The probability $\prod (u)$ of an existing node $u$ to get a new connection depends on its degree $k_u$. It is defined as,

\begin{center}
$\prod (u) = \frac{k_u}{\sum_{v}k_v}$
\end{center}

So, the nodes having higher degrees acquire more links over time, thereby skewing the distribution towards lower degrees. Preferential attachment model gives rise to power law degree distribution where, the probability $f(k)$ of a node having degree $k$ is defined as, 

\begin{center}\begin{equation} \label{eq:fk}
f(k) = ck^{-\gamma}
\end{equation}\end{center}
where, $\gamma$ is the power law exponent, and for real world scale-free networks its range is $2 < \gamma < 3$. As the network grows, only a few nodes called hubs manage to get a large number of links. 

\subsection{Degree Ranking}

In this section, we propose a method to predict degree rank of a node in scale-free networks. Let's consider a scale-free network $G$ having $n$ nodes that follows power law degree distribution $f$. The probability of a node $u$ having degree $k$ is defined as,
\begin{center}
$P(k_u = k) = f(k)$
\end{center}
where, $f(k)$ is power law function from equation \eqref{eq:fk}. 

$f(k)$ contains two parameters $c$ and $\gamma$. First we compute the value of $c$. Using the law of probability, the integration of $f(k)$ from $k_{min}$ to $k_{max}$ will be equal to $1$, where $k_{min}$ and $k_{max}$ are the minimum and maximum degree in the network.
\begin{center}
$\int_{k_{min} }^{k_{max} }f(k) dk = 1 $\\
$\int_{k_{min} }^{k_{max} } c k^{-\gamma} dk = 1 $
\end{center}
\begin{center}
$c \frac{ (k_{min})^{1-\gamma} - (k_{max})^{1-\gamma} }{ \gamma - 1 } = 1$
\end{center}
\begin{center}
$c = \frac{ \gamma - 1 }{ (k_{min})^{1-\gamma} - (k_{max})^{1-\gamma} }$
\end{center}

To compute the value of $\gamma$, we use average degree $d_{avg}$ of the network, that can be written as,

\begin{center}
$d_{avg} = \int_{k_{min} }^{k_{max} } kf(k) dk $
\end{center}
\begin{center}
$d_{avg} = \int_{k_{min} }^{k_{max} } kck^{-\gamma} dk$
\end{center}
After integration,
\begin{center}
$d_{avg} = c \frac{ k_{min}^{2-\gamma} - k_{max}^{2-\gamma}}{\gamma-2} $
\end{center}
Using value of $c$, the equation will be,
\begin{center}
$d_{avg} = \frac{\gamma-1}{\gamma-2} (\frac{ k_{min}^{2-\gamma} - k_{max}^{2-\gamma}}{ k_{min}^{1-\gamma} - k_{max}^{1-\gamma} }) (k_{max}k_{min}) $
\end{center}
\begin{center}
$d_{avg} = \frac{\gamma-1}{\gamma-2} (\frac{ k_{max}^{\gamma-2} - k_{min}^{\gamma-2}}{ k_{max}^{\gamma-1} - k_{min}^{\gamma-1} }) (k_{max}k_{min}) $
\end{center}
where, $k_{min} << k_{max}$ and $2 < \gamma < 3$ for scale-free networks,
\begin{center}
$d_{avg} = \frac{\gamma-1}{\gamma-2} \frac{k_{max}^{\gamma-2}}{k_{max}^{\gamma-1}} (k_{min} k_{max})$
\end{center}
\begin{center}
$d_{avg} = \frac{\gamma-1}{\gamma-2} k_{min}$
\end{center}
\begin{center}
i.e. $ \gamma = 2 + \frac{k_{min}}{d_{avg}-k_{min}}$
\end{center}

Now, the rank of a node $u$ having degree $k_u=k$ can be estimated as,
\begin{center}
$R_{p}(u) = n - n \int_{k_{min}}^{k}f(k) dk +1$\\
$R_{p}(u) = n - n \int_{k_{min}}^{k}ck^{-\gamma} dk +1$
\end{center}
After integration, 

\begin{center}
$R_{p}(u) = n- nc \frac{k^{1-\gamma} - k_{min}^{1-\gamma}}{ {1-\gamma} } +1$\\
$R_{p}(u) = n- nc \frac{k_{min}^{1-\gamma} - k^{1-\gamma}}{ {\gamma-1} } +1$
\end{center}
Using value of $c$, 
\begin{center}
$R_{p}(u) = n- n\frac{k_{min}^{1-\gamma} - k^{1-\gamma}}{ k_{min}^{1-\gamma} - k_{max}^{1-\gamma} } +1$
\end{center}
\begin{center}
$R_{p}(u) = n( \frac{k^{1-\gamma} - k_{max}^{1-\gamma}}{ k_{min}^{1-\gamma} - k_{max}^{1-\gamma} } ) +1$
\end{center}
\begin{center}
$R_{p}(u) = ( \frac{n}{ k_{min}^{1-\gamma} - k_{max}^{1-\gamma} } )k^{1-\gamma} - (\frac{ n k_{max}^{1-\gamma}}{ k_{min}^{1-\gamma} - k_{max}^{1-\gamma} } -1)$
\end{center}
\begin{center}
$R_{p}(u) = ak^{1-\gamma} + b$
\end{center}
where, $a=\frac{n}{ k_{min}^{1-\gamma} - k_{max}^{1-\gamma} }$ and $b=\frac{ n k_{max}^{1-\gamma}}{ k_{min}^{1-\gamma} - k_{max}^{1-\gamma} } -1$. $a$ and $b$ are constants for a given network.

\section{Simulation Results}

\begin{figure}[ht!]
     \begin{center}
        \subfigure[]{%
            \label{fig:first}
            \includegraphics[width=0.5\textwidth]{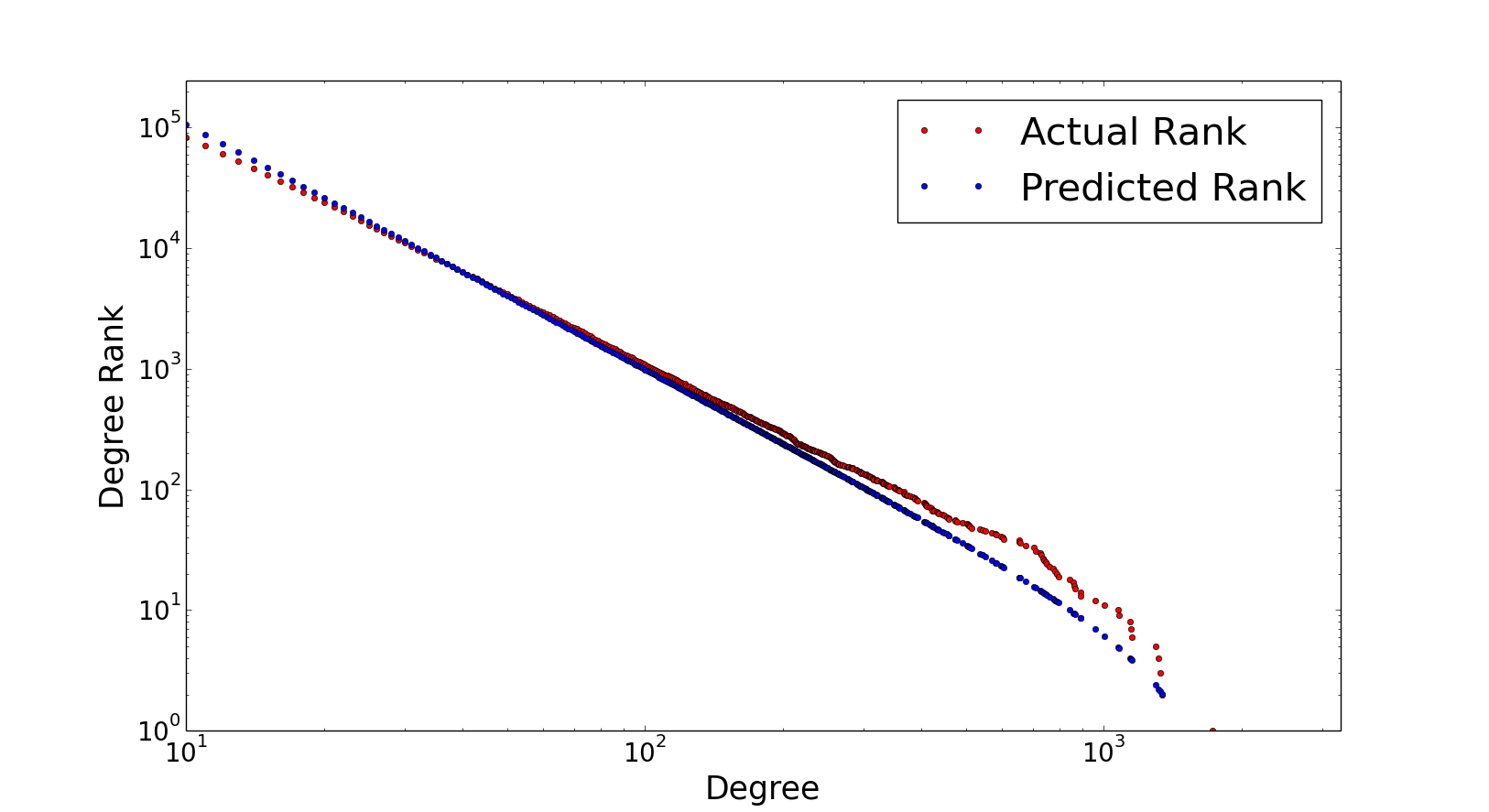}
        }%
        \subfigure[]{%
           \label{fig:second}
           \includegraphics[width=0.5\textwidth]{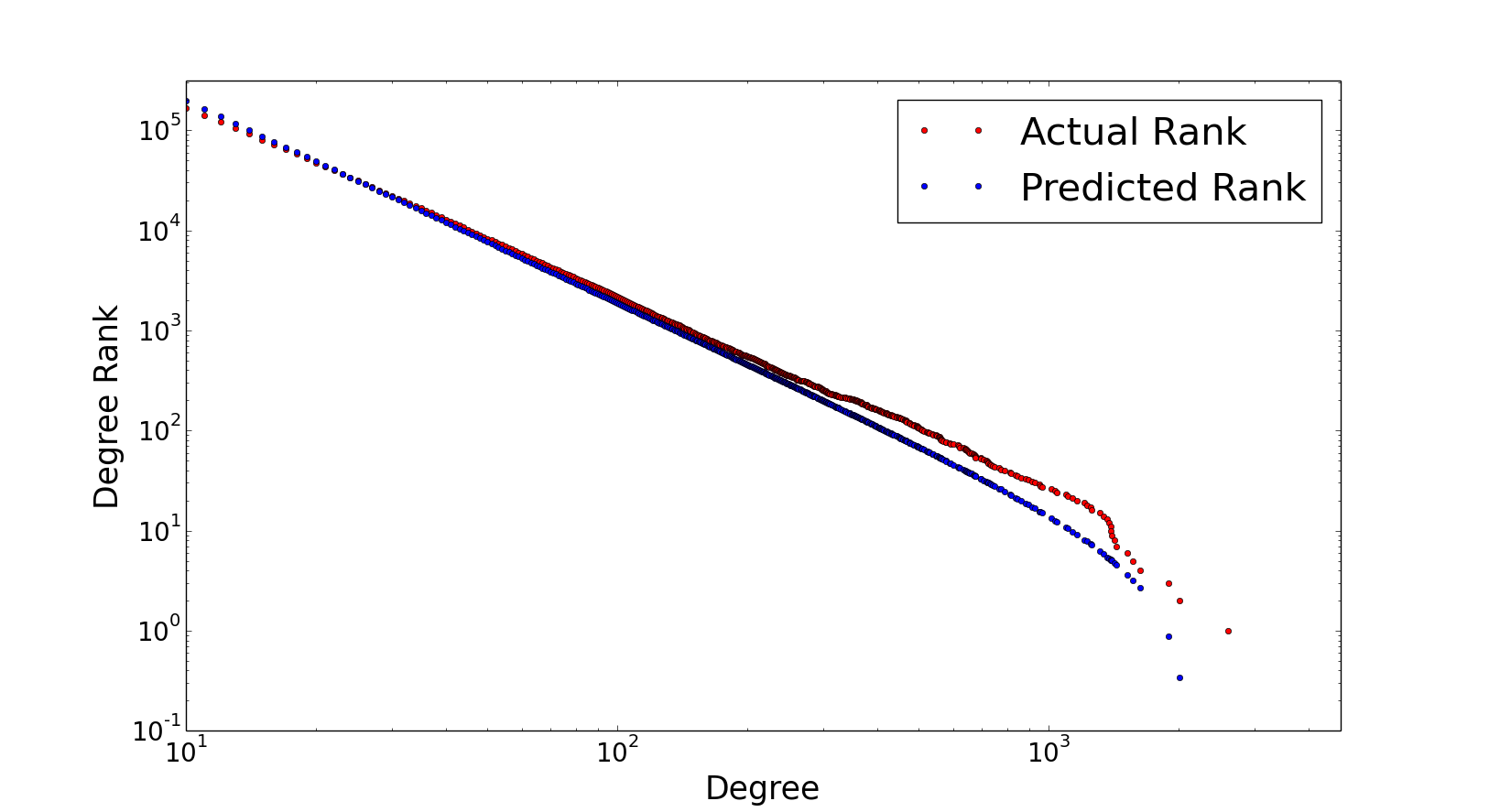}
        }\\ 
        \subfigure[]{%
            \label{fig:third}
            \includegraphics[width=0.5\textwidth]{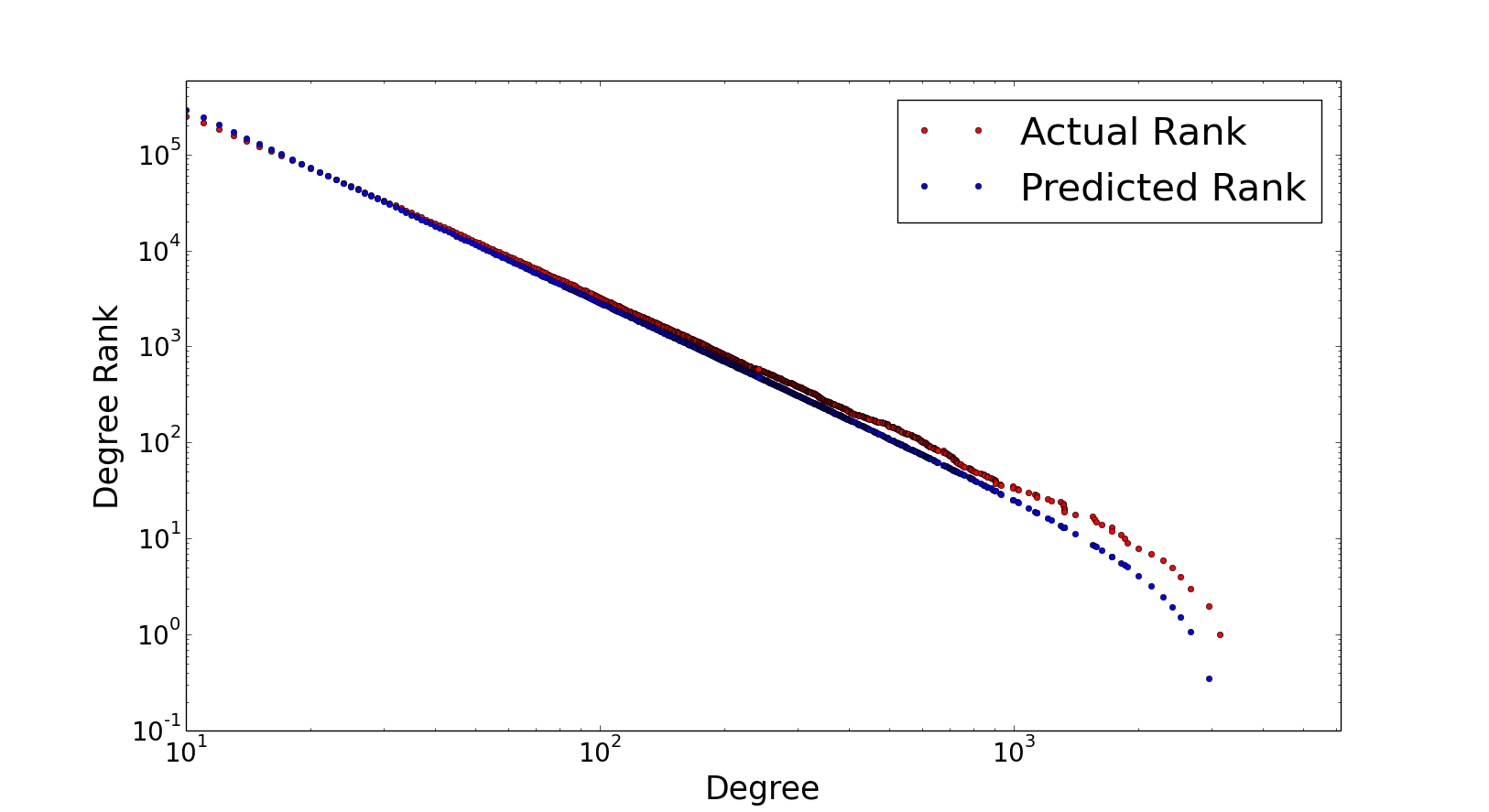}
        }%
        \subfigure[]{%
            \label{fig:fourth}
            \includegraphics[width=0.5\textwidth]{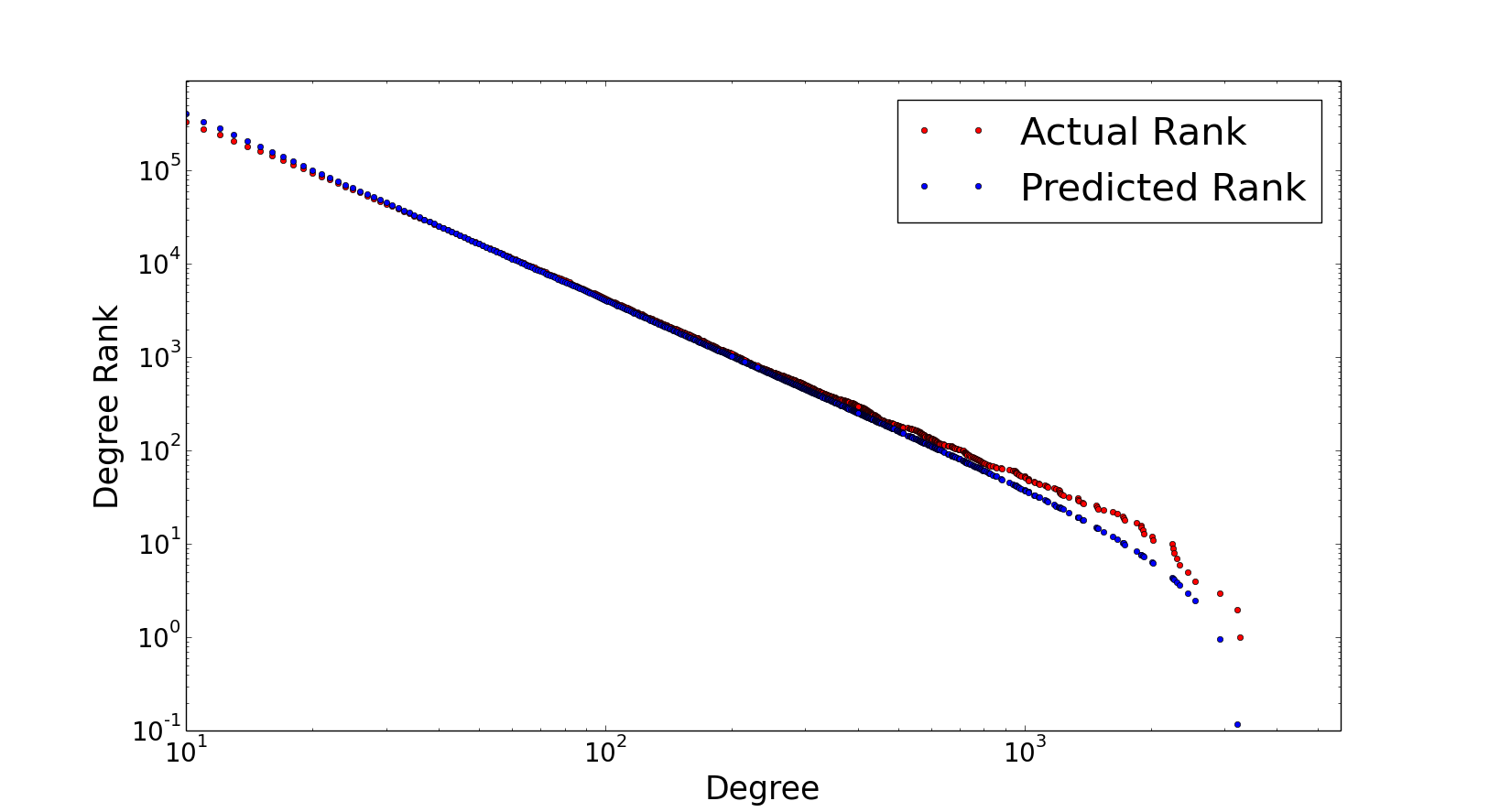}
        }\\ 
          \subfigure[]{%
            \label{fig:fifth}
            \includegraphics[width=0.5\textwidth]{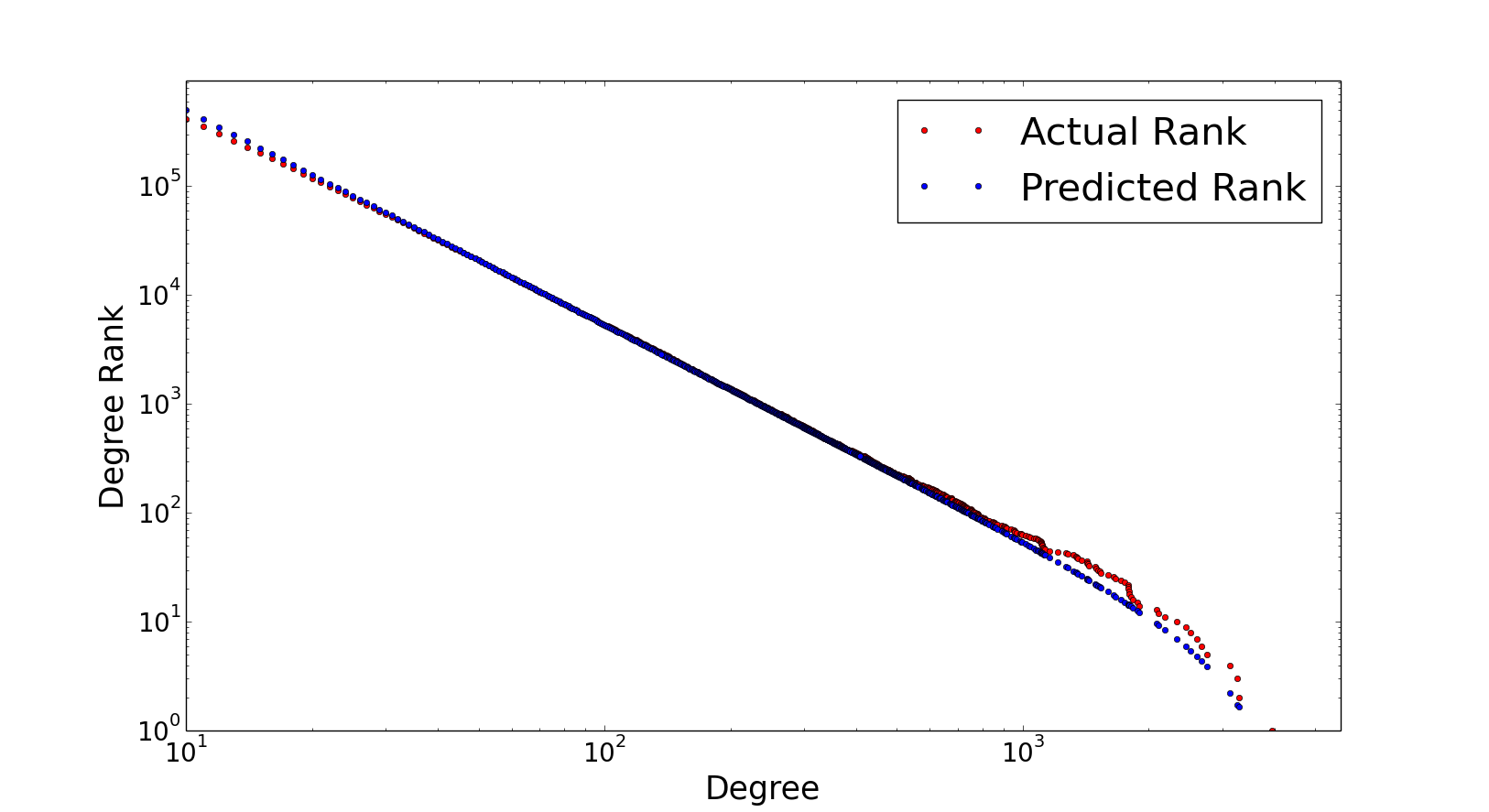}
        }
    \end{center}
    \caption{Plots of actual and estimated rank on log-log scale for BA Network having a)100000 b)200000 c)300000 d)400000 e)500000 nodes.}
   \label{fig:subfigures}
\end{figure}

The proposed method is simulated on BA networks of $100000-500000$ nodes. BA networks are generated with a seed graph having 10 nodes, and each new coming node makes $10$ connections using preferential attachment law. To simulate the proposed method, we first need to estimate the required network parameters. The network size is estimated using the method proposed by Hardiman and Katzir \cite{hardiman2013estimating}, and the average degree is estimated using the method proposed by Dasgupta et al. \cite{dasgupta2014estimating}. The network size estimation method converges when approximately $1\%$ nodes of the network are sampled. The minimum and maximum degree is estimated as the minimum and maximum degree available in the sample. To measure the accuracy of the proposed method, the absolute error is computed for each degree, that is defined as $Absolute\; Error(k)= |Actual\; Rank(k)-Predicted\; Rank(k)|$. The absolute error is averaged over all degrees to compute the overall error for the network.  

In Fig. 1, the actual and the predicted degree rank are plotted for BA networks. According to the method, highest degree node has rank $1$. It can be observed in Fig. 1 that the prediction for higher degree nodes is more accurate than the lower degree nodes. This difference occurs because the power law degree distribution function is integrated from minimum to maximum degree. It assumes that the nodes of all degrees are present in the network, but it might not be true for the generated networks. The one more reason is that the generated networks may not follow perfect power law as per the estimated power law exponent. Table 1 shows the average error and standard deviation of the predicted rank for all networks. The results show that the average error is approximately $0.16\%$ of the network size. The average absolute error increases with the network size, but the percentage average absolute error decreases with the network size. The percentage average absolute error is computed as, (average absolute error *100)/network size.

\begin{table}[]
\centering
\caption{Error in the Actual and Predicted Ranking}
\label{my-label1}
\begin{tabular}{|l|l|l|l|l|}
\hline
Number of Nodes & Average Error & Standard Deviation \\ \hline

100000 & 335.02 & 1833.64 \\ \hline
200000 & 389.63 & 2128.77 \\ \hline
300000 & 447.87 & 2580.50 \\ \hline
400000 & 600.52 & 4531.16 \\ \hline
500000 & 631.58 & 5057.44 \\ \hline
\end{tabular}
\end{table}

\section{Conclusion}

In this work, we have proposed a method to estimate degree centrality rank of a node without having the entire network. The proposed method is based on the power law exponent of degree distribution of the network. The simulation results show that the average error in the predicted rank is around $0.16\%$ of the network size. We will further validate the accuracy of the proposed method on real world scale-free networks. This work can be applied to identify influential nodes in real world networks.

In future, we will extend this work to estimate the lower and upper bound on degree centrality rank of a node. With time, the size of complex networks is increasing very fast. It will be of great help, if local information of the network can be used to estimate the rank of a node based on global centrality measures such as betweenness centrality, closeness centrality, pagerank, and so on.

\bibliographystyle{unsrt}
\bibliography{mybib.bib}

\end{document}